**Using NIST Special Publications (SP) 800-171r2 and 800-172/800-172A to assess and evaluate the Cybersecurity posture of Information Systems in the Healthcare sector[1].**

Thomas P. Dover[2]

**ABSTRACT**

This paper describes how NIST Special Publications (SP) 800-171r2 (*Protecting Controlled but Unclassified Information in Nonfederal Systems and Organizations*), SP.800-172 (*Enhanced Security Requirements for Protecting Controlled Unclassified Information*) and SP.800-172A (*Assessing Enhanced Security Requirements for Controlled Unclassified Information*) can be used to evaluate the cybersecurity posture of information systems and supporting frameworks relative to HIPAA[3] and HITECH[4]. It will demonstrate that provisions and baseline security requirements outlined in SP.800-171r2 and SP.800-172/172A for the protection of Controlled Unclassified Information (CUI) can be applied to Electronic Protected Health Information (ePHI). An explanation of how these publications align with HIPAA and how this alignment suffices for evaluating IT environment security will be given along with the process and procedure for performing such evaluation. Finally, the benefits of using this approach to support formal risk assessment will be presented.

**TARGET AUDIENCE**

Organizations or individuals responsible for the administration, management, operations, security and maintenance of Information Technology systems and applications which hold, transfer or store Electronic Protected Health Information (ePHI). This includes both datacenter and cloud-based service providers.

**Note:** though this paper is specific to healthcare its approach can be applied to any industry or sector.

**SCOPE**

The process and procedures outlined in this paper are directed towards organizations or businesses whose information systems are responsible for providing service to, or otherwise supporting, indirectly or directly, healthcare providers, services, organizations or any other entity which handles, stores or transmits ePHI.

---

[1] This paper (version 3) updates the August, 2020 version to reflect publication of SP.800-172A (*Assessing Enhanced Security Requirements for Controlled Unclassified Information*), published April, 2021.
[2] Adjunct Faculty, Butler County Community College, Butler, PA. Email: thomas.dover@bc3.edu
[3] Health Insurance Portability and Accountability Act (1996)
[4] Health Information Technology for Economic and Clinical Health Act (2009)

**U.S. Critical Infrastructure Sector:** Healthcare and Public Health Sector
**Responsible Department:** Health and Human Services (HHS)

**INTRODUCTION**

In June, 2015 the National Institute of Standards and Technology (NIST) released Special Publication SP.800-171 (*Protecting Controlled Unclassified Information in Nonfederal Systems and Organizations*). This publication was succeeded by SP.800-171r1 in December, 2016 and followed by SP.800-171r2 and its supplement SP.800-171B (*Protecting Controlled Unclassified Information in Nonfederal Systems and Organizations, Enhanced Security Requirements for Critical Programs and High Value Assets*) in June, 2019. The current version of SP.800-171r2 was released in February, 2020. SP.800-171B was renamed SP-800-172 (Draft) and released in July, 2020.

The purpose of SP.800-171r2 is to provide non-federal organizations[5] with guidance for protecting the Confidentiality[6] of unclassified (but controlled) information. As stated in its Abstract

> This publication provides agencies with recommended security requirements for protecting the confidentiality of CUI when the information is resident in nonfederal systems and organizations; when the nonfederal organization is not collecting or maintaining information on behalf of a federal agency or using or operating a system on behalf of an agency; and where there are no specific safeguarding requirements for protecting the confidentiality of CUI prescribed by the authorizing law, regulation, or government-wide policy for the CUI category listed in the CUI Registry. The requirements apply to all components of nonfederal systems and organizations that process, store, or transmit CUI, or that provide security protection for such components.

Though not specifically intended for healthcare organizations, requirements contained within SP.800-171r2 are nevertheless applicable to the healthcare sector[7]. This is due, in part, to the Department of Health and Human Services (HHS) and Office of Civil Rights (OCR) references to NIST publications[8] in published documents and online guidance regarding cybersecurity protections and HIPAA's Security Rule. For example, HHS-OCR, Security Risk Assessment (SRA) Tool[9] references NIST publications in its User Guide. Moreover, vendors who develop security tools (e.g., IDS, anti-virus, email protection) and Managed Security Service Providers (MSSP) often refer to NIST guidelines as "industry-standards".

In April, 2021, NIST released (Draft) SP.800-172A (*Assessing Enhanced Security Requirements for Controlled Unclassified Information*). Per NIST, "This generalized assessment procedures described in this publication provide a framework and starting point for developing specific procedures to assess the enhanced security requirements in NIST Special Publication 800-172[10]."

---

[5] Primarily federal contractors, or companies, agencies or organizations doing business with the federal government.
[6] Confidentiality is one part of the control triad for protecting sensitive information such as Electronic Health Record(s). The other parts of the triad are Integrity and Availability. Together, they form the CIA of cybersecurity.
[7] "…processing…healthcare data;" SP.800-171r2, p1.
[8] Examples include the NIST Cybersecurity Framework and SP.800-53r5 (Control catalog)
[9] Available at The Office of the National Coordinator for Health Information Technology, Office of Civil Rights, Department of Health and Human Services (https://www.healthit.gov/topic/privacy-security-and-hipaa/security-risk-assessment-tool)
[10] See 'Cautionary Note', p. vi.

Copyright© Thomas P. Dover. 2021. All Rights Reserved.2

In effect, SP.800-172A uses *determination statements* and *Organization-defined parameters* as procedures for meeting assessment objectives. Meeting a determination statement results in a finding of *Satisfied* or *Other than Satisfied*. It then introduces three specific assessment methods (*Examine, Interview, Test*[11]) for defining the "nature *and extent of the assessor's actions."* Also introduced are associated attributes *Depth* and *Coverage*.

*Examine* is "the process of checking inspecting reviewing…to facilitate understanding, achieve clarification or obtain evidence."

*Interview* is "the process of conducting discussions with individuals or groups…to facilitate understanding, achieve clarification or lead to the location of evidence."

*Test* is "the process of exercising one or more assessment objects under specific conditions to compare actual with expected behavior."

Each assessment method may contain one or more Object *Specifications*, *Mechanisms*, and *Activities*. These objects serve as evidence or proof through which assessment method requirements were met.

The attributes *Depth* and *Coverage* are used to describe the depth (i.e., rigor) and breadth (i.e., scope) of the assessment method review. For each method one of three values (*Basic*, *Focused* and *Comprehensive*) is used to describe the level of analysis.

*Basic* employs "high-level reviews, checks, observations or inspections of the assessment object."

*Focused* employs the above plus "more in-depth studies and analysis."

*Comprehensive* employs the above plus "detailed, and thorough studies and analyses of the assessment object."

As can be seen, each value represents a greater depth and breadth of analysis & review for a particular assessment method.

It should be noted that in its 'Cautionary Note'[12] statement NIST notes that the assessment methods and objects "do not necessarily reflect, and should not be directly associated with, compliance or noncompliance with the requirements."

An example of how to employ the assessment methods and their attributes of SP.800-172A will be given in the Security Assessment Workbook section of this paper.

---

[11] SP.800-172A, Appendix C provides extensive description and explanation of all assessment methods and attributes.
[12] Ch3, p.7

**THE CHALLENGE OF THE SECURITY RULE**

For healthcare providers the essential requirement of the HIPAA Security Rule is that "Covered Entities[13]" take steps to protect the Confidentiality, Integrity and Availability (CIA) of Electronic Protected Health Information (ePHI) through Administrative, Technical and Physical (ATP) means. ATP controls address the *privacy* of information while CIA concerns itself with the *security* of information.

Most clinical care providers (i.e., hospitals, clinics, medical practices, etc…) are supported by an internal Information Technology (IT) department, a third-party provider, or a combination of both. Whether belonging to a 'Covered Entity' or 'Business Associate' it is the responsibility of the IT Services provider to evaluate its security posture relative to the control, protection and management of ePHI.

**SECURITY EVALUATION AND RISK ANALYSIS**

A requirement of the Security Rule is that Risk Analysis[14] be conducted. Such analysis differs from the security evaluation outlined in SP.800-171r2 and SP.800-172/172A in the following ways:

- A Risk Analysis evaluates Likelihood, Impact and Consequence
- A Risk Analysis evaluates Threat, Vulnerability and Expected Loss

The derived benefit of a security evaluation is that it can serve as an initial (or interim) step towards completion of formal Risk Analysis by pointing to weaknesses in an organization's (IT) security framework. Using a sliding Yes/No scale to rate requirement satisfaction and 0-1 for statistical valuation allows an organization to a) identify weaknesses and strengths and b) gauge overall effectiveness of its security environment using a well-defined set of requirements. Results can then be used to complement or supplement formal Risk Assessments.

In order to evaluate adherence to the Security Rule, healthcare providers must identify a) controls applied to systems containing ePHI and b) assess the degree at which they are performing. Such evaluation can be a difficult and challenging task given today's diversified networks and software applications.

---

[13] and Business Associates (BA)

[14] 45 CFR 164.308(a)(1)(ii)(a); "*Conduct an accurate and thorough assessment of the potential risks and vulnerabilities to the confidentiality, integrity and availability of electronic protected health information held by the covered entity.*"

**NIST SP.800-171r2 and SP.800-172/172A SECURITY EVALUATION PROCESS**

Special Publication 800-171r2 utilizes FIPS-200[15] and SP.800-53[16] as the basis for its recommended Security Requirements.  FIPS-200 defines the minimal security requirements for Low, Medium and High-impact information systems as outlined in FIPS-199[17].   NIST SP.800-53r5[18] identifies twenty (20) 'control families'[19] and SP.800-171r2 utilizes a subset of these families.  Control Families are groupings of security controls which address a specific security requirement.  For example, *Access Control* deals with the methods, processes and/or procedures by which a user is granted access to a network or system.

SP.800-171r2 (and by association SP.800-172/172A) omits[20] seven control families contained in SP.800-53r5 that are specific to the federal government.  It uses the remaining 13 'control families' but also incorporates a single, unique control family (*Security Assessment*).  Together, these 14 control families form the basis for its one hundred ten (110) security-control requirements.

Table 1 displays SP.800-53r5 control families[21].  Those highlighted in gray/bold have been omitted from SP.800-171r2 security baseline requirements due to their unique 'federal' nature.  Tailoring requirements in this manner makes application easier and results more accurate when applied to non-government sectors such as healthcare.

| Access Control | Physical and Environmental Protection[22] |
|---|---|
| **Assessment, Authorization and Monitoring** | **PII Processing and Transparency** |
| Audit and Accountability | **Planning** |
| Awareness and Training | **Program Management** |
| Configuration Management | Risk Assessment |
| **Contingency Planning** | **System and Services Acquisition** |
| Identification and Authentication | *Security Assessment[23] |
| Incident Response | System and Communications Protection |
| Maintenance | System and Information Integrity |
| Media protection | **Supply Chain Risk Management** |
| Personnel Security | |

Table 1

---

[15] Federal Information Processing Standards (FIPS) 200, *Minimum Security Requirements for Federal Information and Information Systems.* NIST. Released March, 2006.
[16] NIST SP.800-53r5, *Security and Privacy Controls for Information Systems and Organizations*. Released August 2017.  Final draft published March, 2020.
[17] Federal Information Processing Standards (FIPS) 199, Standards for Security Categorization of Federal Information and Information Systems, NIST. Released February, 2004.
[18] NIST SP.800-53r5, *Security and Privacy Controls for Information Systems and Organizations*. Released August, 2017.  Final draft released March, 2020.
[19] Control Families are security controls (applied to technology systems) which are operational, technical and management (i.e., administrative) safeguards used to protect the confidentiality, integrity and availability (CIA) of information systems.
[20] "…some of the security requirements expressed in the NIST standards and guidelines are uniquely federal, the requirements in this publication have been tailored for nonfederal agencies." SP.800-171r2, p.3.
[21] Note: *Security Assessment* is not an SP.800-53r5 control family.  It is listed here for reference but is cited only in SP.800-171r2 and SP.800-172.
[22] SP.800-171r2 cites this family as 'Physical Protection'
[23] This control family is not included in SP.800-53r5 but is unique to SP.800-172. Reference only.

**EVALUATION PROCESS**

*The process for completing a* SP.800-171r2 *(medium-level security requirement))* or SP.800-171r2 and SP.800-172/172A *(enhanced/high-level security requirement)* assessment consists of satisfying each control requirement then determining compliance for both individual and aggregate control families.

It should be stressed that SP.800-171r2 defines its control baseline security level as being for **moderate-impact** information systems and such level would cover healthcare-service providers handling, transmitting or storing ePHI. SP.800-172 requirements are *enhancements* to SP.800-171r2 and therefore offer stronger security which would be needed for **high-impact** information systems. SP.800-172 contains thirty-four (34) enhanced security-control requirements and SP.800-172A offers assessment methods for evaluating assurance with SP.800-172 requirements.

An advantage of using both SP.800-171r2 and SP.800-172/172A is that security assessments can be performed from the perspective of both medium and enhanced-level security. Evaluating this way allows an organization to determine the level of compliance for each security level and to what extent it is being implemented.

In addition to control-requirement baseline, SP.800-172 has incorporated a new metric called Adversary Effects. Per NIST, "…*adversary effects*…describe the potential effects of implementing the enhanced security requirements on risk, specifically by reducing the likelihood of threat events, the ability of threat events to cause harm, and the extent of that harm. Five high-level, desired effects on the adversary can be identified: *redirect*, *preclude*, *impede*, *limit*, and *expose*." [8] For Adversary Effects, a simple (aggregate) matrix has been created to view the overall impact of security-control implementation.

**SECURITY ASSESSMENT WORKBOOK**

Microsoft Excel was used to create the Security Assessment Workbook[24]. Control-families were placed in separate worksheets (Figure 1) along with summary Snapshot, Compliance, and Adversary Effects[25] worksheets. Finally, several Informational Reference worksheets were included. Formulas, calculated cells and cell references are extensively employed in order to simplify data entry and avoid input or calculation error.

In addition to security-control requirements all control family worksheets contained the following column/row variables:

- Satisfaction of Requirement (Y/N/P/A/D)
- Value (0-1)
- Satisfying Statement (maps to SP.800-172A *Examine* assessment method)
- Name (maps to SP.800-172A *Interview* assessment method)
- Validation Point/Tool (text) {maps to SP.800-172A *Test* assessment method)
- Security Control-HIPAA Type (Administrative, Technical, Physical)

---

[24] Note: the Workbook is not include as part of this paper but can be obtained by contacting the author.
[25] Snapshot, Compliance and Adversary Map worksheets display aggregate data pulled from control-family worksheets.

SP.800-172 (only) Enhanced Security Requirements contain the following column/row variables:

- Assessment Methodology (Examine, Interview, Test)
- Depth & Coverage (Basic, Focused, Comprehensive)

Values for each column/variable is as follows:

**SATISFACTION with REQUIREMENT**
The organization:

(**Y**) performs this task[26]
(**P**) partially performs this task
(**A**) uses an alternate approach to perform this task (that satisfies the requirement)
(**N**) does not perform this task
(**D**) this control requirement does not apply

**VALUE**
Depending on compliance, the following values are assigned:
**Y** or **A** = 1
**P** = Low (.25), Medium (.5), High (.75)
**N** or **D** = 0

**SATISFYING STATEMENT**
Normally, security assessments require detailed explanations[27] of policies and/or procedures in order to satisfy a particular security requirement. Such detail often requires a large allocation of resources (i.e., time and staff) to complete. The approach taken here is to provide a "trimmed" answer but one which satisfies the requirement. For example, in ACCESS CONTROL, the question (do you) "Limit system access to authorized users, processes acting on behalf of authorized users, and devices (including other systems)". A simple, yet satisfactory response would be (yes via) "Role-based Access Control (RBAC)". The trimmed statement is placed in the cell directly below the control-requirement.

**NAME**
Person or persons who were contacted and validated the security requirement.

**VALIDATION POINT/TOOL**
This information denotes what specific application, utility, or process is used to satisfy the control requirement. Often, the same tool is used to satisfy several requirements or several tools for the same requirement. (example would be an IDS[28] and VPN for remote access security).

---

[26] The term 'task' implies steps taken to satisfy the control requirement.
[27] The level of detail need only be sufficient enough to satisfy the security requirement. Nevertheless, more often than not extensive and detailed explanations are given.
[28] Intrusion Detection System (IDS)

## SECURITY CONTROL/TYPE

This value directly references HIPAA's Security Rule which requires security controls to be categorized as *Administrative*, *Technical* or *Physical*. Most often, a security control has but a single categorization but there are instances where a control may encompass more than one. For example, establishing an operational incident-handling capability may be categorized as both an Administrative and Technical control.

## COMPLETION AND COMPLIANCE

For data entry, only control-requirement satisfaction is needed for computation; answers are automatically translated into numerical values and both are used to compute satisfaction. Satisfaction Statement, Name, Validation Point/Tool, HIPAA Security Control (Type) and Assessment Method values for Depth and Coverage are entered manually (Figure 1). Information from each control-requirement worksheet is then used as input for Snapshot and Compliance worksheets (**note:** Adversary Map data is manually entered).

Figure 1

A closer view of incorporation of SP.800-172A assessment methods is provided in Figure 2. Note: each SP.800-172 requirement would include the below table/matrix (as seen in Figure 1).

Figure 2

In the context of a security assessment, *completion* is the extent to which all questions have been answered (i.e., either they have or haven't). Given the number of questions in the assessment[29] the more complete an assessment is the more accurate the results.

*Compliance* is defined as the extent an organization's security 'posture' is aligned with and satisfies individual requirements of SP.800-171r2 (Medium-security) or SP.800-171r2 and SP.800-172/172A (Enhanced/High-security). Compliance establishes whether or not, for a given security-control the Validation Tool maps to the requirement. This process is especially useful in identifying security requirements with either no associated tool or an insufficient one.

Point value and compliance percentage is computed for each control-requirement worksheet with results displayed at the bottom of each sheet (see Figure 2). As stated earlier, this information is used as reference for the Snapshot worksheet and as input for the Compliance worksheet. The Snapshot worksheet provides an aggregate view of all Control-Family responses as well as a summary of completion and compliance (Figure 3).

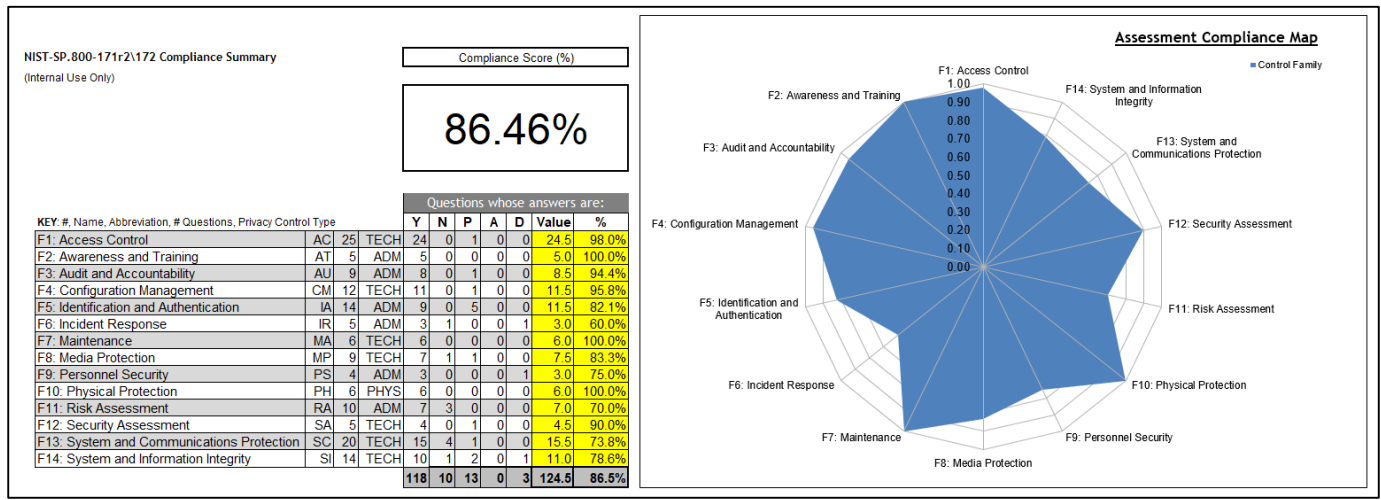

Figure 3 (**note:** highlighted cell used for all non-yes responses for readability and emphasis)

Compliance is summarized via the Compliance Summary worksheet (Figure 4)

Figure 4

---

[29] SP.800-171r2 contains 110 control requirements and SP.800-172 has 34. Both publications comprise a total of 144 security-control requirements.

A data table and radar[30] (aka spider) chart provide tabular and graphical depiction of each Control Family's value for aggregate compliance. The radar chart is especially useful for viewing deficiencies and areas which need to be addressed.

An acceptable compliance level is left to the discretion of the organization as there is no published standard (although 80% or better would be generally accepted). Acceptable levels, however can be designated for both control-family and aggregate levels.

Regardless of threshold, Compliance provides an organization with an idea of how well its security posture compares to established or recommended industry-standards.

Finally, Adversary Effects have been included in SP.800-172 but since they are not ordinal it is not possible to quantify their value. As such, a simple map has been created to show which effects have been achieved if an enhanced security-control is implemented (Figure 5).

**NIST-SP.800-172 Adversary Effects Map**
(Ref. Appendix D.)

| Control Family | # | C | (R)edirect | (P)reclude | (I)mpede | (L)imit | (E)xpose |
|---|---|---|---|---|---|---|---|
| F1: Access Control | 23 | Y | | Yes | Yes | | |
| | 24 | Y | | Yes | Yes | | |
| | 25 | Y | | Yes | Yes | | |
| F2: Awareness and Training | 4 | Y | | | Yes | | Yes |
| | 5 | N | | | No | | No |
| F3: Audit and Accountability | | | No Adverse Effects Listed | | | | |
| F4: Configuration Management | 10 | N | | | No | No | No |
| | 11 | P | | Yes | Yes | | Yes |
| | 12 | Y | | | | | Yes |
| F5: Identification and Authentication | 12 | N | No | | | | No |
| | 13 | Y | | | Yes | | |
| | 14 | Y | | Yes | | | Yes |
| F6: Incident Response | 4 | D | | | | No | No |
| | 5 | Y | | Yes | Yes | Yes | Yes |
| F7: Maintenance | | | No Adverse Effects Listed | | | | |
| F8: Media Protection | | | No Adverse Effects Listed | | | | |
| F9: Personnel Security | 3 | Y | | Yes | Yes | | |
| | 4 | N | | | | No | |
| F10: Physical Protection | | | No Adverse Effects Listed | | | | |
| F11: Risk Assessment | 4 | Y | | Yes | Yes | | Yes |
| | 5 | Y | | Yes | | Yes | Yes |
| | 6 | N | No Adverse Effects Listed | | | | |
| | 7 | Y | No Adverse Effects Listed | | | | |
| | 8 | N | | | | | No |
| | 9 | Y | | Yes | | | Yes |
| | 10 | Y | | Yes | Yes | | |
| F12: Security Assessment | 5 | Y | | | Yes | | Yes |

Figure 5

---

[30] A radar chart compares the values of three or more variables relative to a central point. It's useful when you cannot directly compare the variables and is especially great for visualizing performance analysis or survey data.

**BENEFITS OF USING NIST SP.800-171r2 and SP.800-172/172A FOR SECURITY EVALUATION**

1. SP.800-171r2 and SP. 800-172/172A can be used to assess an organization's information security posture in alignment with HIPAA requirements.
2. The derived security requirements outlined in NIST SP.800-171r2 and SP.800-172/172A omit security controls exclusive to federal information systems.
3. Utilizing the assessment approach described in SP.800-171r2 and SP.800-172/172A suggests an evaluation methodology that is adaptable to the non-federal sector.
4. Areas of weakness identified in a security assessment can be used, in part, as the basis for formal risk assessment.
5. The security assessment approach offered by SP.800-171r2 and SP.800-172/172A is easy to use, flexible, repeatable, and employs industry Best-Practices and guidance.

**CONCLUSION**

Healthcare providers (i.e., Covered Entities) are mandated by HIPAA and HITECH to protect the Confidentiality, Integrity and Availability (CIA) of Electronic Protected Health Information (ePHI). This requirement extends to technology providers and Business Associates (BA) who directly support (healthcare) providers.

NIST SP.800-171r2 and SP.800-172/172A can be used to evaluate an organization's security posture relative to data protection. Moreover, each provides a different level of security for evaluation; SP.800-171r2 for Medium-security systems and SP.800-172 where Enhanced (High) security is required. Since the approach of these publications differs from other NIST publications (in that the emphasis is on protecting information in non-federal information systems) it can be used by non-federal organizations and sectors.

This paper has demonstrated how NIST publications can be used to evaluate a healthcare or IT provider's information security environment by assessing the level of Satisfaction using control-family requirements listed in SP.800-171r2 and SP.800-172. In the aggregate, control-family compliance provides "situational awareness" for how well an organization's security posture aligns with (its) industry standards and best practices.

Finally, SP.800-171r2 and SP.900-172/172A can be used to create a documented history of compliance. Such history can serve as a guide for an organization when considering changes to its information security environment. It can also be used to support regulatory or legal compliance.